\documentclass[amssymb,amsfonts,aps]{revtex4}

\def\limfunc#1{\mathop{\rm #1}}
\def\func#1{\mathop{\rm #1}\nolimits}

\def\ii{\'{\i}}

\def\aa{\`{a}}

\def\cc{\c{c}}

\begin{document}

\title{Phases of quantum states in completely positive non-unitary 
evolution \footnote{This work was partially supported by Funda{\cc}\~{a}o
para o Amparo {\aa} Pesquisa do Estado de S\~{a}o Paulo, FAPESP, and
Conselho Nacional de Desenvolvimento e Pesquisa, CNPq.}}
\author{J. G. Peixoto de Faria \thanks{Electronic address:
jgfaria@fma.if.usp.br}} 
\author {A. F. R. de Toledo Piza} 
\address{Departamento de F{\ii}sica-Matem\'{a}tica,
Instituto de F{\ii}sica, \\
Universidade de S\~{a}o Paulo, C.P. 66318, 05315-970, \\
S\~{a}o Paulo, S\~{a}o Paulo, Brazil}
\author{M. C. Nemes}
\address{Departamento de F{\ii}sica, Instituto de Ci\^{e}ncias Exatas, \\
Universidade Federal de Minas Gerais, C.P. 702, 30161-970, \\
Belo Horizonte, Minas Gerais, Brazil}
\date{\today}

\begin{abstract}
We define an \textit{operational} notion of phases in interferometry 
for a quantum system
undergoing a completely positive non-unitary evolution. This definition
is based on the concepts of quantum measurement
theory. The suitable generalization of the Pancharatnan connection
allows us to determine the dynamical and geometrical parts of the total
phase between two states linked by a completely positive map.
These results reduce to the known expressions of the total, dynamical
and geometrical phases for pure and mixed states evolving unitarily.
\end{abstract}

\maketitle

\vskip 2 truecm 

\noindent
\textit{Introduction: }
The interest in the holonomy concept has considerably
grown in physics in the last years. Such activity began with the publication 
of Berry's work \cite{berry}. Berry showed that the energy eigenfunctions
accumalated a phase factor of purely geometrical origin during an adiabatic
cyclic evolution. Simon \cite{simon} gave a mathematical interpretation
of Berry's phase as the holonomy of a complex line bundle. Later,
Wilczek and Zee \cite{wz} generalized this concept to non-abelian
phases associated with cyclic evolution of degenerate hamiltonians in
the adiabatic limit; Aharonov and Anandan \cite{aa} showed that the 
adiabatic limit is unnecessary and Samuel and Bhandari 
\cite{samuel-bhandari}, using the ideas developed by Pancharatnam
\cite{pancha}, showed that the Berry's phase arises even in 
noncyclic and non-unitary evolutions (see also Ref. \cite{mukunda-simon}). 
These treatments of the 
geometric phases are restricted only to pure states, like the 
majority of the works on this topic. However, in a large number
of applicatons (in particular, the Quantum Theory of Measurement 
\cite{QTM} and the Theory of Quantum Open Systems 
\cite{quantum-open-systems}), 
the concept of state must be generalized
to include statistical mixtures. The question of the mixed state 
holonomy was already studied by Uhlmann \cite{uhlmann} in a purely 
mathematical context.  Sj\"{o}qvist and co-workers \cite{sjoqvist}
recently developed an \textit{operational} concept of phases for
mixed states undergoing unitary evolution using an interferometric
model. Despite these achievements, the study of geometric phases
remains still incipient particularly for mixed states undergoing 
\textit{non-unitary} evolution, characteristic of quantum open
systems. 

In the present work we construct an \textit{operational} definition
of phases between mixed states undergoing \textit{completely
positive} (CP) non-unitary 
time evolution. We employed a procedure similar to that used by 
Sj\"{o}qvist \textit{et al.} \cite{sjoqvist} in the context of
unitarily evolving mixed states in 
interferometry. Moreover, by virtue of the close ties between  
CP non-unitary evolution and continuous generalized measurement processes,
we used some results of the Quantum Measurement Theory
\cite{non-unit}. The suitable
generalization of the Pancharatnam connection allows us to identify 
the dynamical and geometrical contents of the total phase between
two states under CP non-unitary evolution. The results obtained
reduce to the corresponding expressions for total, geometrical and
dynamical phases for unitarily evolving mixed states. 
\\

\noindent
\textit{Completely positive mapping:  }
A mapping $\Phi$ between density operators $\rho$ and $\rho'$ is
said to be completely positive
(CP)  \cite{quantum-open-systems,sandu,gorini,lind,stinespring} 
iff it has the form $\rho'=\Phi(\rho)=\sum_{n}W_{n}\rho W^{\dag}_{n}$,
where $\left\{ W_{n} \right\}$ are operators acting on the Hilbert
space of the system of interest. If $\Phi$ preserves the trace, the
set  $\left\{ W_{n} \right\}$ constitutes a partition of the unity:
\begin{equation}
\sum_{n}W_{n}^{\dag} W_{n} = \mathbf{1} 
\label{POVM} 
\end{equation}
A CP continuous time evolution $\Phi_t$ is governed by the dynamical
generator \cite{gorini,lind}
\begin{equation}
\dot{\rho}=\frac{1}{i\hbar}\left[ H, \rho\right]+\frac{1}{2}
\sum_{n}\left(2 V_{n}\rho V^{\dag}_{n} - \rho V^{\dag}_{n}
V_{n} - V^{\dag}_{n}V_{n} \rho \right),
\label{gen}
\end{equation}
where $H$ is a hamiltonian operator and $\left\{ V_{n}\right\}$ is
a set of operators on the Hilbert space of the system of interest.
Eq. (\ref{gen}) is called the Lindblad equation.

A CP non-unitary continuous time evolution describes a \textit{
non-selective} or \textit{non-referring continuous measurement process}. 
In other words, at each infinitesimal time interval  $dt$, the system 
suffers a generalized measurement and its state, after  $dt$, is
given by 
\[
\rho\left( t+dt\right)=\sum_{n}W_{n}\left( dt\right)\rho\left( 
t\right)W^{\dag}_{n}\left( dt\right),
\]
where the \textit{measurement operators} $\left\{ W_{n}\left( dt\right)
\right\}$ are 
\begin{eqnarray}
W_{0}\left( dt\right)&=&1-\left( \frac{i}{\hbar}H+\frac{1}{2}\sum_{n >
0}V^{\dag}_{n}V_{n} \right) dt, \nonumber \\
W_{n}\left( dt\right)&=&V_{n}\sqrt{dt},\: n = 1,2,\ldots. 
\label{meas-ops}
\end{eqnarray}
In a non-selective or non-referring measurement, the corresponding
outcome is ignored.
Each measurement operator is associeted to one of the possible 
measurement outcomes. The null result, in particular, is associated
to the operator $W_{0}\left( dt\right)$. A null outcome occurs when
no ``pointer'' deflection or no ``click'' is recorded by the meter. In other
words, a null outcome occurs when the meter state is not changed 
during the measurement process. 
\\

\noindent
\textit{Phases and measurements in interferometry: }
Consider a conventional Mach-Zhender interferometer, as shown in
Fig. 1. A beam of particles follows through the paths labelled as 
``0'' and ``1''. These paths define a two dimensional Hilbert space
$\tilde{\mathcal{H}}=\left\{ \left| \tilde{0} \right>,
\left| \tilde{1} \right> \right\}$. The state vectors 
$\left| \tilde{0} \right>$ and $\left| \tilde{1} \right>$ represent
wave packets moving along the arms of the interferometer. The action of
the mirrors, beam-splitters and the phase $U(1)$ shift 
due the difference of paths are represented by the unitary 
operators
\[\tilde{U}_{M}=\left( 
\begin{array}{cc}
  0 & i \\
  i & 0 \\
\end{array}
\right), \:
\tilde{U}_{B}=\frac{1}{\sqrt{2}}\left( 
\begin{array}{cc}
  1 & i \\
  i & 1 \\
\end{array}
\right),\: 
\tilde{U}(1)=\left( 
\begin{array}{cc}
  e^{i\chi} & 0 \\
  0         & 1 \\
\end{array}
\right).
\]

As shown in the Fig. 1, the beam enters in the ``0'' arm. The 
``incoming'' state is $\tilde{\rho}_{in}=|\tilde{0}\rangle \langle
\tilde{0} |$. The beam reaches the detector $D$ in the 
``outcoming'' state $\tilde{\rho}_{out}=\tilde{U}_{B}\tilde{U}_{M}
\tilde{U}(1) \tilde{U}_{B}\tilde{\rho}_{in}\tilde{U}_{B}^{\dagger}
\tilde{U}^{\dagger}(1)\tilde{U}_{M}^{\dagger}\tilde{U}_{B}^{\dagger}$.
The intensity measured by the detector is proportional to the matrix
element  $\langle\tilde{0} | \tilde{\rho}_{out} |\tilde{0}\rangle $, i.e.,
$I\propto 1+\cos \chi$. 

Following the formalism of Sj\"{o}qvist \textit{et al.} \cite{sjoqvist}, 
let us assume that the particles have internal degrees of freedom
(e.g., spin or polarization). The corresponding $N$-dimensional state 
space is denoted 
by ${\mathcal{H}}_{i}$. The initial state of the internal degrees of
freedom is represented by the operator
$\rho_{i,0}$. Moreover, let us assume that the mirrors and beam-splitters
do not influence the internal state. 
In order to obtain the phase relation between states under CP non-unitary 
evolution, a meter $P$ is introduced into the ``1'' arm of the
interferometer, as shown in Fig. 1. The meter performs measurements on
the internal degrees of freedom of the particles that cross the interferometer.
By virtue of the inclusion of the meter, we are compelled to consider
a third system associated with the meter -- the probe, 
whose state space is denoted by 
${\mathcal{H}}_{p}$.

The evolution of the composed system particles + internal degrees
of freedom + probe after the first beam-splitter is given by the
operator 
\begin{equation}
{\mathbf{U}}=\left(
\begin{array}{cc}
0 & 0 \\
0 & 1 \\
\end{array}
\right) \otimes U_{ip} + \left(
\begin{array}{cc}
  e^{i\chi} & 0 \\
  0         & 0 \\
\end{array}
\right) \otimes {\mathbf{1}}_{i}\otimes {\mathbf{1}}_{p},
\label{non-unit-ev}
\end{equation}
where $U_{ip}$ is an unitary operator acting in the product space 
${\mathcal{H}}_{i}\otimes{\mathcal{H}}_{p}$. This operator describes the
pre-measurement stage. The operators assigned to the other devices
of the interferometer are ${\mathbf{U}}_{B}=\tilde{U}_{B}\otimes
{\mathbf{1}}_{i} \otimes {\mathbf{1}}_{p},\: {\mathbf{U}}_{M}=
\tilde{U}_{M}\otimes {\mathbf{1}}_{i}\otimes {\mathbf{1}}_{p}$.
Here, ${\mathbf{1}}_{i}$ and ${\mathbf{1}}_{p}$ are the identity
operators on ${\mathcal{H}}_{i}$ and ${\mathcal{H}}_{p}$,
respectively.

Without loss of generality, we assume that the probe is 
prepared in the pure state $|\zeta_{0}\rangle$.
The state of the system particles + internal degrees
of freedom + probe when the beam enters in the interferometer is
thus $|\tilde{0} \rangle \langle \tilde{0}|
\otimes \rho_{i,0}\otimes |\zeta_{0}\rangle \langle \zeta_{0}|$.
After the pre-measurement stage, the state of the global system is 
projected onto a vector belonging to a suitably chosen ``pointer'' basis
of ${\mathcal{H}}_{p}$, namely $\left\{ \left| \zeta_{n} 
\right> \right\}_{n = 0,1,\ldots}$. 
The intensity measured by the detector depends on the result of the
measurement. In the case of a non-null result, we have
\[
I_{n} \propto \func{tr}\left( W^{\dag}_{n}W_{n}\rho_{i,0} \right), \:
W_{n} \equiv \left< \zeta_{n} \right| U_{ip}
\left| \zeta_{0} \right>, \: n = 1, 2, \ldots,
\]
i.e., the \textit{interference pattern vanishes}. In fact, a non-null
result turns the paths distinguishable and 
destroys the interference pattern. If a null result is 
obtained, the intensity measured by the detector is 
\[
I_{0} \propto 1+\func{tr}\left( W^{\dag}_{0}W_{0}\rho_{i,0} \right)
+ 2\nu_{0}\cos\left( \chi - \phi_{0} \right), \:
W_{0} \equiv \left< \zeta_{0} \right| U_{ip}
\left| \zeta_{0} \right>,
\]
where we have defined the visibility $\nu_{0}=\left| \func{tr}\left(W_{0}
\rho_{i,0}\right) \right|$
and the phase shift $\phi_{0}=\func{arg} \func{tr}\left(W_{0}
\rho_{i,0}\right)$.
Therefore, only the null outcome produces non-vanishing visibility 
and contributes to form an interference pattern.

Finally, in the case of a non-selective measurement the intensity 
measured is given by the sum of the intensities $I_{n}$:
\begin{equation}
I = \sum_{n}I_{n} \propto 1+\nu_{0}\cos\left(\chi-\phi_{0}
\right).
\label{tot-int}
\end{equation}
\\

\noindent
\textit{The limit of continuous measurements: }
In order to introduce time into the non-unitary dynamics, we assume 
that the meter
$P$ performs repeatedly generalized measurements on the internal degrees of 
freedom. Each measurement takes a time interval $\Delta t$. 
Hence, in a unique measurement, the internal degrees of freedom
unitarily interact with the probe and this interaction is described
by the operator $U_{ip}\left( \Delta t\right) $. After this stage,
the state of the global system particles + internal degrees
of freedom + probe is projected onto a state of the basis  
$\left\{ \left| \zeta_{n} \right> \right\}_{n=0,1,\ldots}$ chosen
in ${\mathcal{H}}_{p}$. Finally, the state of the probe is 
restored to the ``quiescent'' state (reset of the meter) and the 
process is repeated. The limit of continuous measurement
is obtained taking $\Delta t \rightarrow 0$ and 
$N \rightarrow \infty$, $N\Delta t= t$.

Assuming that the initial state of the internal degrees
of freedom is $\rho_{i}\left( 0\right) $ and the quiescent state 
of the probe is $\left| \zeta_{0}\right> $,  the beam
reaches the detector $D$ in the state
\begin{eqnarray}
\rho_{out}\left(t \right) &=&\frac{1}{4} \left\{ \left(
\begin{array}{cc}
1  & i \\
-i & 1 \\
\end{array}
\right) \otimes \rho_{i}\left( 0\right) 
+ \left(
\begin{array}{cc}
e^{i\chi}   & -ie^{i\chi} \\
-ie^{i\chi} & -e^{i\chi} \\
\end{array}
\right) \otimes \left[ \rho_{i}\left( 0\right)
S^{\dagger}\left(t\right) \right] \right. \nonumber \\
& &\left. + \left(
\begin{array}{cc}
e^{-i\chi}  & ie^{-i\chi} \\
ie^{-i\chi} & -e^{-i\chi} \\
\end{array}
\right) \otimes \left[ S\left(t\right)
\rho_{i}\left( 0\right)  \right] 
+ \left(
\begin{array}{cc}
1 & -i \\
i & 1 \\
\end{array}
\right) \otimes \rho_{i}\left( t\right)  \right\}  ,
\label{rho-out-t}
\end{eqnarray}
where we have defined
\begin{equation}
S\left(t\right) = \exp \left[-\left(\frac{i}{\hbar}
H + \frac{1}{2}\sum_{n > 0} V_{n}^{\dagger} V_{n}
\right) t\right] .
\label{S-t}
\end{equation}
It is worthwhile to remark that the evolution of the density operator
$\rho_{i} \left( t\right)$ obeys the Lindblad equation (\ref{gen}).

The intensity measured by the detector $D$ is 
\[
I \propto 1 + \nu \cos \left[\chi-\phi\left(0,t\right)\right],
\]
where we have defined the visibility of the interference pattern
\begin{equation}
\nu=\left|\limfunc{tr}\left[S\left( t\right) \rho_{i}\left(0
\right) \right]\right|
\label{visibility-t}
\end{equation}
and the total phase difference between the states $\rho_{i}\left(0\right)$ 
and $\rho_{i}\left(t\right)$
\begin{equation}
\phi\left(0,t\right)=\arg \limfunc{tr}\left[S\left( t\right) 
\rho_{i}\left( 0 \right) \right].
\label{tot-ph}
\end{equation}
\\

\noindent
\textit{Parallel transport: }
Let us suppose that a quantum system initially prepared in the 
state  $\rho_{0}=\rho\left( 0\right)$ evolves non-unitarily during 
a time interval $t$. This evolution is CP and the dynamical generator 
has the form (\ref{gen}). The total phase between $\rho \left( t\right)$
and $\rho_{0}$ is
\begin{equation}
\phi \left(0,t\right)=\func{arg}\func{tr} \left[S\left( t\right) 
\rho_{0} \right].
\label{non-un-global-ph}
\end{equation}

We can employ the parallel transport condition 
to obtain the purely
geometrical contribution to the total phase $\phi$
\cite{reference-section}. In fact, the
phase between the states $\rho \left( t+dt\right)$ and 
$\rho \left( t\right)$ is $\phi \left( dt\right)=\func{arg}\func{tr}
\left[ W_{0}\left( dt\right) \rho \left( t\right) \right]$,
where $W_{0}\left( dt\right)$ is given by (\ref{meas-ops}). Hence,
\[
\phi \left( dt\right)= \func{arg} \left\{ 1- \frac{dt}{2}\sum_{n>0}
\func{tr}\left[
V^{\dag}_{n}V_{n} \rho \left( t\right)\right]+\frac{dt}{i\hbar}
\func{tr}\left[ H \rho\left( t\right) \right] \right\}.
\]
$\rho \left( t+dt\right)$ and $\rho \left( t\right)$ are in phase
if $\phi\left( dt\right)=0$. Note that 
\[
1- \frac{dt}{2}\sum_{n>0}\func{tr}\left[
V^{\dag}_{n}V_{n} \rho \left( t\right)\right] \geq 0,
\]
since
\[
0 \leq \sum_{n>0} V^{\dag}_{n}V_{n} dt \leq \sum_{n}W^{\dag}_{n}
\left( dt\right) W_{n}\left( dt\right) = \mathbf{1}.
\]
Therefore, in order to satisfy the parallel transport condition 
it is necessary that
\begin{equation}
\func{tr}\left[ H\rho \left( t\right) \right] = 0.
\label{par-transp-3}
\end{equation}

This result is the analogue of the condition (12) of \cite{sjoqvist} for the 
parallel transport of mixed states under unitary evolution. Hence, if a state
suffers a parallel displacement along a CP non-unitary curve
$\Gamma: \tau \in \left[ 0,t \right] \rightarrow \rho
\left( \tau \right)$ between the extreme points $\rho \left( 0\right)$ and
$\rho \left( t\right)$ such that Eq. (\ref{par-transp-3}) is
satisfied, the total phase $\phi\left(0, t\right)$ has  a purely geometrical
character, i.e., $\gamma_{g}\left(\Gamma \right) = \phi\left(0, t\right)$.
However, the condition stated in above equation is not sufficient.
In the context of CP maps, a sufficient criteria is given by 
Eq. (20) of Ref. \cite{ericsson}.
We can define the dynamical phase as  
\begin{equation}
\gamma_{d}\left(0,t \right) =-\frac{1}{\hbar}\int_{0}^{t}d\tau 
\func{tr}\left[\rho\left( \tau \right)H \right].
\label{dyn-ph-2}
\end{equation}

Eq. (\ref{non-un-global-ph}) constitutes the generalization of 
the Pancharatnam connection for quantum states undergoing a CP 
non-unitary evolution. In fact, in the unitary limit, the Eqs. 
(\ref{non-un-global-ph}-\ref{dyn-ph-2})
reduce to the expressions obtained by Sj\"{o}qvist and co-workers.
\\

\noindent
\textit{Example:  }
Let us consider a single mode of the electromagnetic field undergoing
a continuous and destructive photo-counting process. The evolution
of the field state $\rho$ is governed by the master equation
\begin{equation}
\dot{\rho}=-i\omega \left[a^{\dagger}a, \rho \right] + k 
\left( 2a\rho a^{\dagger
}-a^{\dagger }a\rho -\rho a^{\dagger }a\right) .  \label{ex7}
\end{equation}
The constant $k$ represents the photo-counting rate per time unit and
$\omega$ stands for the mode frequency. Comparing the master equation 
(\ref{ex7}) with the general form of the quantum dynamical generator
(\ref{gen}), we have
\begin{equation}
V_{1}=\sqrt{2k}a \label{V1}
\end{equation}
and 
\begin{equation}
H=\hbar \omega a^{\dagger }a.  \label{hamilt}
\end{equation}
The master equation (\ref{ex7}) describes the time evolution of a damped
harmonic oscillator at zero temperature. If the mode is prepared in 
the initial coherent state $\rho \left( 0\right) =\left| \alpha 
\right\rangle \left\langle \alpha\right| $ 
it is easy to verify that, after a time interval $t$, the state of
the mode will be
$\rho \left( t\right) =\left| \alpha e^{-\left( i\omega +k\right)
t}\right\rangle \left\langle \alpha e^{-\left( i\omega +k\right) t}\right|$ .
According to Eq. (\ref{tot-ph}), the total phase between $\rho \left(
0\right) $ and $\rho \left( t\right) $ is 
\begin{equation}
\phi \left( 0,t\right) =-\left| \alpha \right| ^{2}e^{-kt}\sin \omega t,
\label{ex10}
\end{equation}
and the dynamical phase is
\begin{equation}
\gamma _{d}\left( 0,t\right) =-\frac{\omega }{2k}\left| \alpha \right|
^{2}\left( 1-e^{-2kt}\right) .  \label{ex11}
\end{equation}
The geometric phase is given by the expression
\begin{equation}
\gamma _{g}\left( 0,t\right) =-\left| \alpha \right| ^{2}\left[ e^{-kt}\sin
\omega t-\frac{\omega }{2k}\left( 1-e^{-2kt}\right) \right] .  \label{ex12}
\end{equation}

The master equation (\ref{ex7}) remains \textit{invariant} under the 
transformations 
\begin{eqnarray}
H^{\prime } &\rightarrow &H-\frac{i\hbar }{2}\left( \beta ^{\ast
}V_{1}-\beta V_{1}^{\dagger }\right) ,  \label{ex13} \\
V_{1}^{\prime } &\rightarrow &V_{1}+\beta ,  \nonumber
\end{eqnarray}
The new measurement operators $W_{0}^{\prime }\left( dt\right) $ and 
$W_{1}^{\prime }\left( dt\right) $ describe a continuous homodyne
photo-counting process \cite{carmichael}. The c-number $\beta$ is assigned
to the amplitude of the local oscillator. Surprisingly, 
$\gamma _{g}^{\prime }\left( 0,t\right) =\gamma _{g}\left(
0,t\right) $. This result  suggests that the geometric content of the phase
between two states of a quantum system is invariant under 
\textit{unitary} transformations of the measurement operators, 
since the non-selective evolution is not affected \cite{qmeas}.

It is worthwhile to note that the phase space of the set of
coherent states of the harmonic oscillator coincides with the
complex plane. An initial coherent state remains coherent during
the time evolution generated by Eq. (\ref{ex7}), although its 
amplitude is damped.
The geometric phase calculated in the two cases (conventional
and homodyne measurements) is proportional to the area
in the complex plane limited by the trajectory described by the
time-dependent amplitude $\alpha\left( \tau\right) = 
\alpha e^{-\left(i\omega + k\right)\tau}$ and the shortest geodesic
linking the extreme points, as shown in the Fig. 2.
\\

\noindent
\textit{Conclusion:  } 
We have obtained an extension of the notion of phases in interferometry
for mixed states under CP non-unitary evolution. We have determined
the dynamical and geometrical parts of the total phase between states
undergoing a CP non-unitary evolution. Our results reduce to the
corresponding expressions of Sj\"{o}qvist \textit{et al.}
\cite{sjoqvist} in the unitary limit.

\newpage
\noindent
\large Figure captions
\\

\normalsize
\noindent
Fig. 1. A conventional Mach-Zhender interferometer. $M_1$ and $M_2$ 
are  perfect mirrors, $B_1$ and $B_2$ are balanced beam-splitters. The
detector $D$ measures the intensity of the beam of particles in ``0'' 
arm. $\chi$ is the $U(1)$ phase shift due the path difference. In
order to introduce non-unitary evolution, a meter $P$ is connected 
into ``1'' arm.
\\

\noindent
Fig. 2. The geometric phase between the coherent states $\rho\left( 0\right)=
\left|\alpha \right>\left< \alpha\right| $ and $\rho\left( t\right)=
\left|\alpha e^{-\left(i\omega + k\right)t}\right>
\left< \alpha e^{-\left(i\omega + k\right)t}\right| $ of the
harmonic oscillator evolving according to the master equation
(\ref{ex7}) is proportional to the area of the region in complex plane 
limited by the trajectory described by the time-dependent amplitude
$\alpha\left( \tau\right) = \alpha e^{-\left(i\omega + k\right)\tau} $,
$\tau \in \left[0,t\right]$, and the shortest geodesic linking the 
extreme points. 

\end{document}